\documentclass[aps,prb,twocolumn,superscriptaddress]{revtex4-1}
\usepackage{graphicx}

\begin{document}

\title{Spin Gap and the Nature of the 4d$^3$ Magnetic Ground State \\in the Frustrated FCC  Antiferromagnet  Ba$_2$YRuO$_6$}

\author{J.~P.~Carlo}
\affiliation{Department of Physics, Villanova University, Villanova, PA  19085  USA}

\author{J.~P.~Clancy}
\affiliation{Department of Physics, University of Toronto, Toronto, ON  Canada}

\author{K.~{Fritsch}}
\affiliation{Department of Physics and Astronomy, McMaster University, Hamilton, ON  L8S 4M1  Canada}

\author{C.~A.~Marjerrison}
\affiliation{Department of Physics and Astronomy, McMaster University, Hamilton, ON  L8S 4M1  Canada}

\author{G.~E.~Granroth}
\affiliation{Quantum Condensed Matter Division, Oak Ridge National Laboratory, Oak Ridge, TN 37831  USA}

\author{J.~E.~Greedan}
\affiliation{Department of Chemistry, McMaster University, Hamilton, ON L8S 4M1 Canada}
\affiliation{Brockhouse Institute for Materials Research, McMaster University, Hamilton, ON L8S 4M1 Canada}

\author{H.~A.~Dabkowska}
\affiliation{Brockhouse Institute for Materials Research, McMaster University, Hamilton, ON L8S 4M1 Canada}

\author{B.~D.~Gaulin}
\affiliation{Department of Physics and Astronomy, McMaster University, Hamilton, ON  L8S 4M1  Canada}
\affiliation{Brockhouse Institute for Materials Research, McMaster University, Hamilton, ON L8S 4M1 Canada}
\affiliation{Canadian Institute for Advanced Research, Toronto, ON M5G 1Z8 Canada}

\date{\today}

\begin{abstract}
The geometrically frustrated double perovskite Ba$_2$YRuO$_6$ has magnetic $4d^3$ Ru$^{5+}$ ions decorating an undistorted face-centered cubic (FCC) lattice. This material has been previously reported to exhibit commensurate long-range antiferromagnetic order below $T_{N} \sim$~36~K, a factor $f \sim 15$ times lower than its Curie-Weiss temperature $\Theta_{CW} = -522$~K, and purported short-range order to $T^*$~=~47~K.    We report new time-of-flight neutron spectroscopy of Ba$_2$YRuO$_6$ which shows the development of a $\sim$5 meV spin gap in the vicinity of the [100] magnetic ordering wavevector below $T_N$~=36~K, with the transition to long-range order occurring at $T^*$ = 47~K.  We also report spin waves extending to $\sim$14 meV, a surprisingly small bandwidth in light of the large $\Theta_{CW}$.  We compare the spin gap and bandwidth to relevant neutron studies of the isostructural 4$d^1$ material Ba$_2$YMoO$_6$, and discuss the results in the framework of relatively strong spin-orbit coupling expected in 4$d$ magnetic systems.

\end{abstract}

%\keywords{}

\maketitle

%\subsection{}
%\subsubsection{}

\section{INTRODUCTION}

Geometrically frustrated magnetic materials \cite{lacroix, gardner_pyrochlores_review} are of great current interest due to the exotic ground states they possess, a consequence of the intrinsic competition between their interactions and anisotropies on appropriate crystalline architectures.   These states include spin liquid,\cite{herbertsmithite_spin_liquid} spin glass, \cite{RamirezPRL2000, GardnerPRL1999YMoO} and spin ice states,\cite{HarrisPRL1998, ClancyPRB, RossPRX} as well as long-range ordered states which form via exotic mechanisms, such as order by disorder.\cite{champion_ErTiO, gaulin_ErTiO, ZhitomirskyPRL,  SavaryRossPRL}  Many such materials are based on two-dimensional (2D) assemblies of triangles and three-dimensional (3D) assemblies of tetrahedra.  In 2D, networks of edge-sharing triangles are common, and triangular magnets such as NaCrO$_2$ \cite{NaCrO2} and VCl$_2$ \cite{VCl2} have been well studied, while organic triangular materials such as $\kappa$-(BEDT-TTF)$_2$Cu$_2$(CN)$_3$ \cite{KanodaPRL91_107001_2003} are of great topical interest.  Kagome nets formed by 2D networks of corner-sharing triangles have also attracted considerable attention;\cite{Wills_PRB61_6156_2000, MatanPRL96_247201_2006} one such s=$^1$/$_2$ system, Herbertsmithite (ZnCu$_3$(OH)$_6$Cl$_2$), appears a likely candidate for a quantum spin liquid state.\cite{herbertsmithite_spin_liquid}

The tetrahedron is to 3D what the triangle is to 2D, and networks of corner-sharing tetrahedra are found and are well-studied in the cubic pyrochlores,\cite{gardner_pyrochlores_review} spinels, and certain Laves phase compounds.  Networks of edge-sharing tetrahedra form the face centered cubic (FCC) lattice.  Despite the fact that the FCC lattice is a dense stacking of triangular layers, and therefore also common in nature, magnetic materials exhibiting this structure with promising indicators of geometrical frustration are relatively uncommon, and have not been as well studied.

The A$_2$BB'O$_6$ double perovskites with magnetic B' cations can form such a FCC magnetic lattice, provided that the B and B' ions are sufficiently distinct to exist in the ``rock-salt"-ordered region of the double perovskite phase diagram\cite{anderson_dblperovskites} (Fig. 1(a)).  The Ba$_2$YB'O$_6$ family, where B' is a magnetic 4$d$ or 5$d$ transition metal element in its $5+$ oxidation state, is very interesting in this regard.  Ba$_2$YMoO$_6$ and Ba$_2$YRuO$_6$ represent examples of 4$d^1$ and 4$d^3$ moments which are antiferromagnetically coupled on an undistorted FCC lattice, respectively.  Related 4$d$ double perovskites, such as Sr$_2$YRuO$_6$,\cite{granado_sr2yruo6} La$_2$LiMoO$_6$\cite{tomoko_1_2} and La$_2$LiRuO$_6$\cite{tomoko_3_2} also exist, but these undergo structural distortions such that the symmetry of their lattices is lower than cubic at low temperatures.  Among 5$d$ double perovskites, Ba$_2$YWO$_6$\cite{kamata_BYWO} and Ba$_2$YReO$_6$\cite{tomoko_1} represent 5$d^1$ and 5$d^2$ moments decorating a FCC lattice, but these systems have received relatively little attention, and their properties are not well understood.

Such 4$d$ and 5$d$ magnetic double perovskites offer the possibility of combining the effects of geometrical frustration with strong spin-orbit coupling (SOC). SOC grows roughly as $Z^4$, and should be appreciably stronger in 4$d$ and 5$d$ systems, relative to their more familiar 3$d$ analogues.  Theoretical studies have indicated rich phase diagrams and exotic ground states in materials combining strong spin-orbit coupling and geometrical frustration.\cite{balents_s_1_2, balents_s_1}

Recent measurements on the undistorted FCC $4d^1$ system Ba$_2$YMoO$_6$\cite{tomoko_1_2} have been interpreted in terms of a spin-liquid, collective singlet ground state.  This system shows a strong antiferromagnetic (AF) Curie-Weiss susceptibility, with $\Theta_{CW}$ = $-$219~K.  Inelastic neutron scattering measurements\cite{carlo_BYMO} found a gapped spin excitation spectrum with a large,  likely singlet-triplet gap of $\sim$28~meV, and weak in-gap states which may be due to weak impurities or structural disorder.  The gap evolves rapidly with increasing temperature, collapsing at $\sim$125~K.  These results are consistent with low-temperature magnetic susceptibility and NMR measurements, both of which suggest a low-temperature phase characterized by the co-existence of a gapped, singlet-like state and a weak paramagnetic state, the latter presumably induced by weak disorder.  Valence bond glass behavior has been proposed in Ba$_2$YMoO$_6$,\cite{deVries_BYMO, deVries_BYMO2} as well as in isostructural Ba$_2$LuMoO$_6$.\cite{Coomer_Cussen_JPhysCondensMatter082202_2013}

The undistorted $4d^3$ analogue, Ba$_2$YRuO$_6$, offers an intriguing comparison.  In the absence of strong SOC, we would expect an orbitally-quenched s=$^3$/$_2$, spin-only moment at the Ru$^{5+}$ site, which should minimize anisotropy and any corresponding spin gap.   It was reported that the temperature dependence of its magnetic susceptibility is characterized by a large and AF $\Theta_{CW}$ = $-$522 K, and that long-range AF order sets in by $T_N$ = 36 K, with another transition possible at $T^*$ = 47 K.\cite{tomoko_3_2}  Prior neutron diffraction measurements \cite{battle_BYRO} revealed Type I commensurate AF order with an effective magnetic moment of $\sim$2~$\mu_B$/Ru$^{5+}$ below $T_N$, although Ba$_2$YRuO$_6$ has $f$ = $\Theta_{CW}$/$T_N$ $\sim$ 15, indicating strong suppression of its ordered state by geometrical frustration, quantum fluctuations, or both.

In this article, we report neutron scattering results on polycrystalline Ba$_2$YRuO$_6$.   We find low-temperature [100] and [110] magnetic Bragg peaks consistent with long-range Type~I AF order, persisting up to $T^*$, the previously-reported short-range ordering temperature.  In addition, strong inelastic magnetic scattering rises from the vicinity of the [100] peak at all temperatures.    However, below $T_N$~$\sim$~36~K an unexpectedly large $\sim$5~meV gap opens up, with the full bandwidth of the spin excitations extending up to about 14~meV.    

\section{ EXPERIMENTAL DETAILS}

Our 10g powder sample of Ba$_2$YRuO$_6$ was prepared by conventional solid-state reaction via the method of Aharen \textit{et al.}\cite{tomoko_3_2}  A stoichiometric mixture of BaCO$_3$, Y$_2$O$_3$ and RuO$_2$ was fired at 1350$^{\circ}$C for a total of 5 days with intermediate re-grindings.  Phase purity was verified with x-ray diffraction, and magnetic susceptibility (Fig. 1(b)) was measured in a field of 500~G.   Curie-Weiss fitting of the inverse susceptibility (inset) revealed an effective moment size $\mu_{\mathrm{eff}}$~=~3.65(1) $\mu_B$ (in comparison to the s=$^3/_2$ spin-only value 3.87 $\mu_B$) and $\Theta_{CW}$ = $-$399(2)~K, consistent with strong and frustrated AF correlations ($f$~$\sim$~11), similar to the results of Aharen \textit{et al.}\cite{tomoko_3_2}

 \begin{figure}[ht]
\includegraphics[width=90mm]{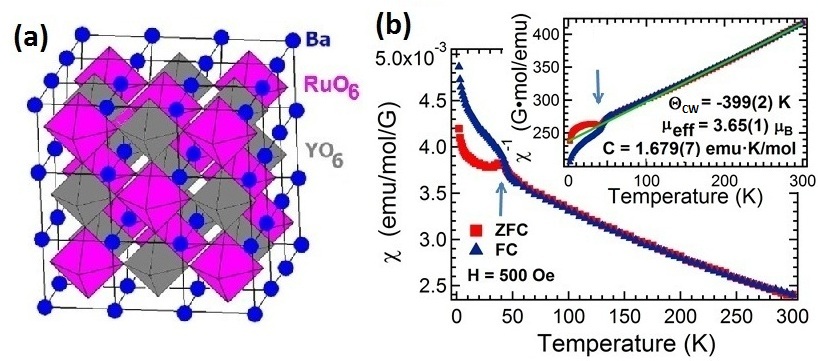}
 \caption{\label{}(a) Unit cell of Ba$_2$YRuO$_6$, with Ru$^{5+}$ ions decorating a magnetic sublattice of edge-sharing tetrahedra. (b) Magnetic susceptibility of Ba$_2$YRuO$_6$, with an arrow indicating $T_N$~=~36~K; from fits to this inverse susceptibility (inset) $\Theta_{CW}$ is found to be $-$399~K.}
 \end{figure}

Neutron scattering measurements were performed at the $SEQUOIA$ Fine Resolution Fermi Chopper Spectrometer at the Spallation Neutron Source (SNS), Oak Ridge National Laboratory.\cite{granroth1, granroth2}   The loose powder specimen was contained in a 5.0~$\times$~5.0~cm (2~mm thick) planar aluminum can in the presence of He exchange gas, and loaded into a closed-cycle refrigerator with a temperature range of 6 K to 290 K.    Time-of-flight measurements employed incident neutron beam energies of $E_i$ =  11~meV chosen by Fermi chopper \#2 spinning at 180~Hz ($\Delta$$E$/$E\sim$5\%), and $E_i$ = 120 meV chosen by Fermi chopper \#1 at 300 Hz.  Background from the prompt pulse was removed by the $T_0$ chopper spinning at 60~Hz (11~meV) or 180~Hz (120~meV).  The sample can was masked with boron nitride to match the sample size, and normalization to a white-beam vanadium run was performed to correct for the detector efficiencies.   An identical empty aluminum sample can was run under the same experimental conditions and used for background subtraction.

\section{NEUTRON SCATTERING RESULTS}

Elastic neutron scattering results at $E_i$ = 11~meV and $E_i$ = 120~meV both show [100] and [110] magnetic Bragg peaks near $|Q|$~=~0.76~$\mathrm{\AA}^{-1}$ and 1.06~$\mathrm{\AA}^{-1}$ for T $\lesssim$ 45~K, consistent with Type I AF order.  $E_i$~=~11~meV elastic scattering data, integrating between $\pm$ 1 meV near the [100] and [110] peaks, are shown as a function of temperature in Fig. 2(a).  Corresponding low-energy inelastic scattering data, integrated between 1 and 2~meV, is shown in Fig. 2(b) as a function of temperature. Two features are clear:  the low-energy inelastic scattering falls off strongly with decreasing temperature, and it extends $\sim$0.3 $\mathrm{\AA}^{-1}$ in $|Q|$ between the [100] and [110] positions.  Were such an asymmetric lineshape to appear in elastic scattering, it could be interpreted in terms of a Warren lineshape, characteristic of two-dimensional correlations within a three-dimensional powder diffraction experiment.\cite{Warren}  Indeed, recent powder neutron diffraction measurements on Sr$_2$YRuO$_6$, without energy discrimination,\cite{granado_sr2yruo6} have reported such a lineshape for intermediate temperatures $T_{N1}=24 $ K $< T < T_{N2}=32$ K (using the nomenclature of Granado \textit{et al}).  In the inelastic spectrum, its interpretation is more subtle, as it originates from the powder-averaged dispersion of the spin excitations in the appropriate energy window.  In Ba$_2$YRuO$_6$, we observe this asymmetric scattering within the inelastic channel only as shown in Fig. 2(b); the elastic Bragg scattering at 45~K, between $T^*$ and $T_N$, shows a resolution-limited, symmetric, Gaussian lineshape, with a roughly $T$-independent width indicating that 3-D long-range order persists up to $T^*$.

 \begin{figure}[ht]
\includegraphics[width=90mm]{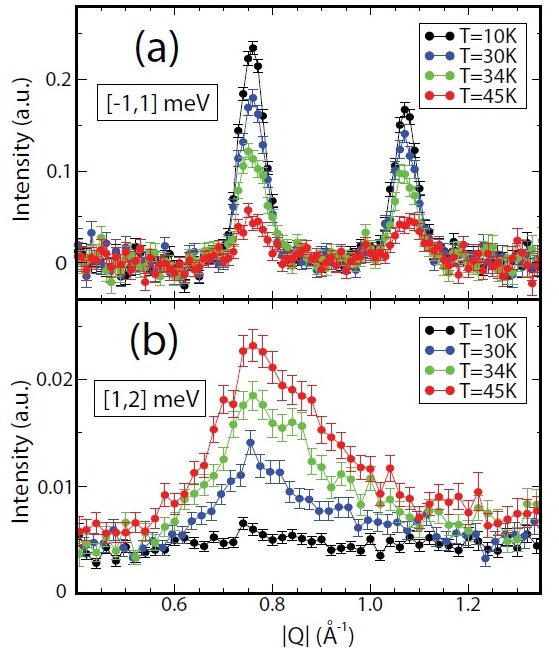}
 \caption{\label{}(a) Temperature dependence of the elastic neutron scattering intensity in Ba$_2$YRuO$_6$, for $E_i$~=~11~meV, integrated over an energy range $\pm$1~meV.  Magnetic Bragg peaks can be observed at the [100] (0.76~$\mathrm{\AA}^{-1}$) and [110] (1.06~$\mathrm{\AA}^{-1}$) wavevectors.   A high temperature ($T$ = 100 K) background has been subtracted from each data set to isolate the magnetic scattering contribution.   (b) Temperature dependence of the inelastic neutron scattering intensity for $E_i$~=~11~meV, integrated over an energy range 1~$<$~$E$~$<$~2~meV.   The low-lying inelastic magnetic scattering is suppressed below $T_N$ = 36 K by the development of a $\Delta$ $\sim$ 5 meV gap, which can be seen in Fig. 3. }
 \end{figure}

\begin{figure}[ht]
 \includegraphics[width=85mm]{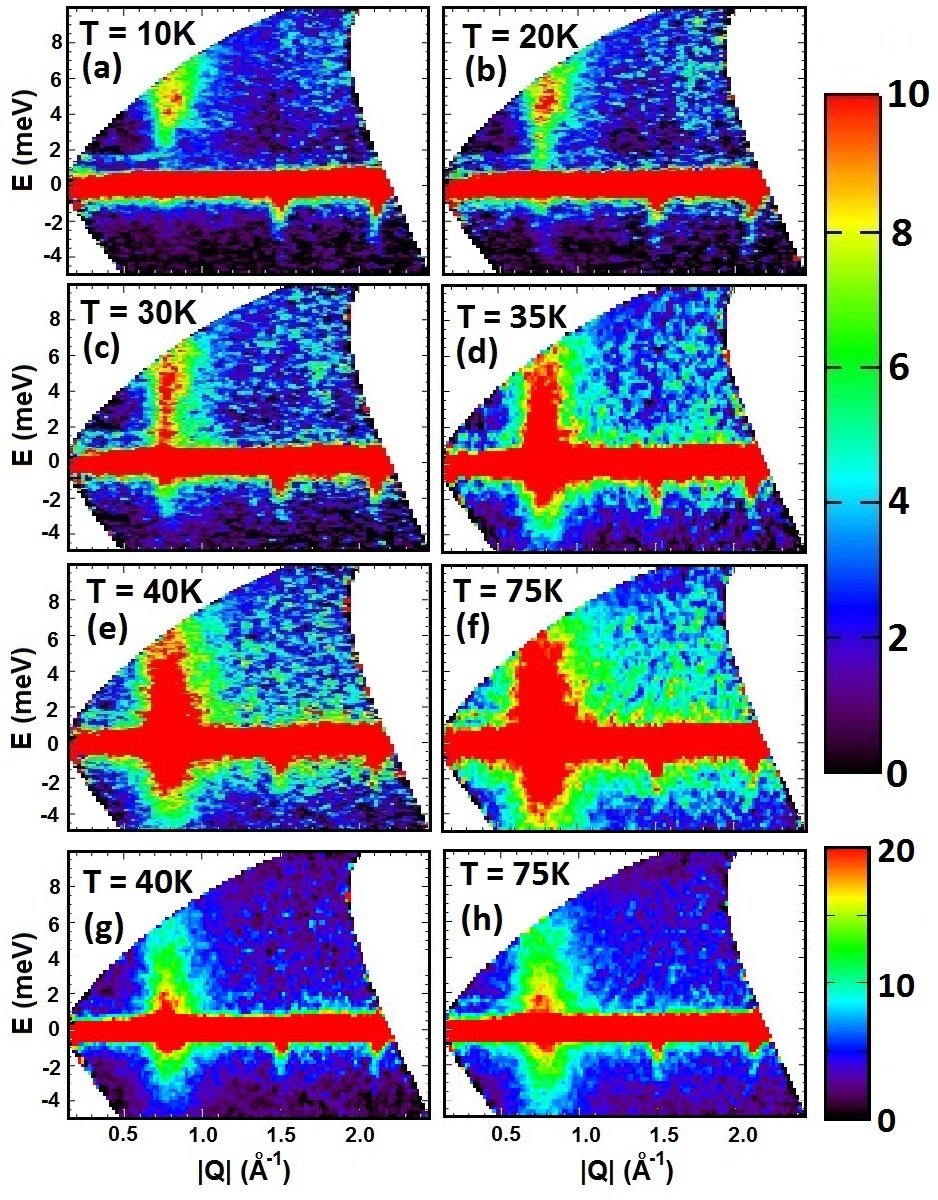}
 \caption{\label{}(a)-(h) Background-subtracted neutron scattering data for Ba$_2$RuYO$_6$ collected with $E_i$ = 11 meV.   The [100] magnetic Bragg peak is located at $|Q|$ = 0.76 $\mathrm{\AA^{-1}}$.    As the temperature drops below $T_N$~=~36~K, a $\sim$~5~meV gap opens.   The horizontal band near 1.5 meV is present in the empty-can runs and is not completely removed by the background subtraction.  The upper intensity scale refers to panels (a)-(f), while the lower intensity scale refers to panels (g) and (h), showing the same temperatures as panels (e) and (f), but with a higher intensity scaling.}
 \end{figure}

 \begin{figure}[ht]
 \includegraphics[width=85mm]{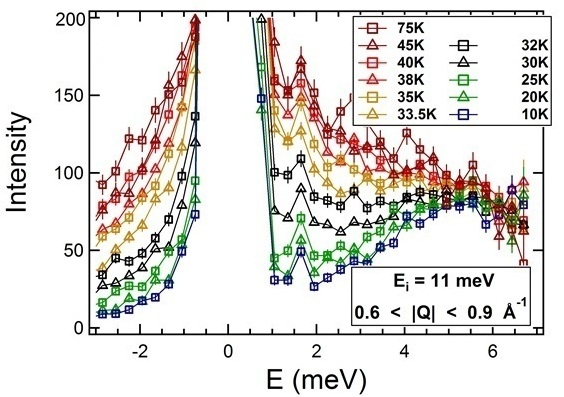}
 \caption{\label{} (a)  Scattering intensity versus energy, integrated over 0.6 $< |Q| < $ 0.9 $\mathrm{\AA}^{-1}$.   The existence of a spin gap below 5~meV is clear with temperature decreasing below $T_N$.}
\end{figure}

Fig. 3 shows our full $E$ vs $|Q|$ neutron data set collected with $E_i$ = 11~meV at temperatures both below and above $T_N$ = 36~K, with the empty-can background subtracted from each data set.   At all temperatures, inelastic scattering rises up from the location of the [100] magnetic Bragg peak at $|Q|$ = 0.76~$\mathrm{\AA}^{-1}$.   A clear gap, of $\sim$5 meV, is seen to form in the inelastic scattering at low temperatures, and it is fully formed by $\sim$20~K, consistent with the results shown in Fig. 2(b).  Fig. 4 shows the energy dependence of the $|Q|$-integrated scattering around the [100] position (0.6~$<$~$|Q|$~$<$ 0.9~$\mathrm{\AA}^{-1}$), clearly showing the opening of the gap in the spin excitation spectrum from the top of the elastic channel up to $\sim$5~meV, and its evolution to a quasielastic spectrum for  $T > T_N$.

One may expect a column of inelastic scattering to exist above both the [110] and [100] ordering wavevectors at low temperatures, but the inelastic scattering is only easily observable at the [100] wavevector well below $T_N$.  This is due to the fact that the [110] elastic  peak is $\sim$30$\%$ weaker than that at [100], and $\sqrt{2}$ further out in $|Q|$, which strengthens the effects of the powder averaging.  As the gap fills in near $T_N$, inelastic scattering becomes stronger due to the expected $1/E$ dependence of $\chi"$, and the Bose factor, which is strong for all $E$~$<$~$\Delta$~$\sim$~$T_N$.  

 \begin{figure}[ht]
 \includegraphics[width=85mm]{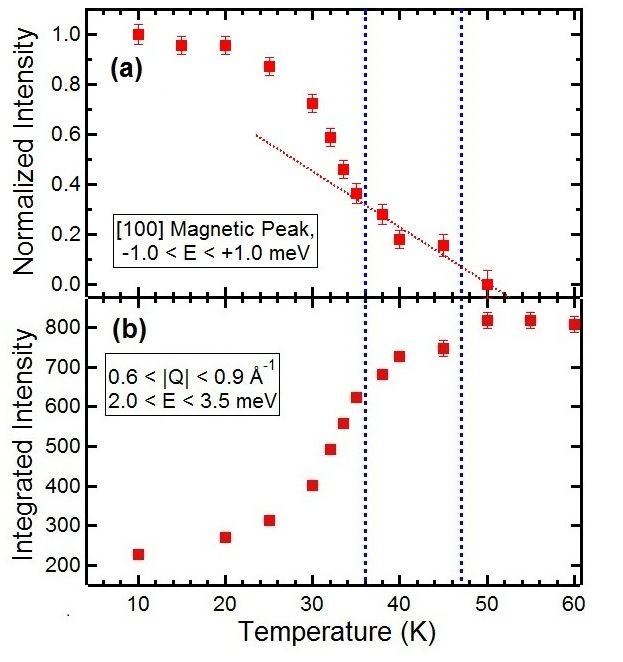}
 \caption{\label{}(a)  Order parameter measurement of the $\pm$1~meV elastic scattering, taken from the data in Fig. 2(a).  Two regimes of temperature dependence to the order parameter are clear: one below $T_N$ and one between $T^*$~=~47~K and $T_N$~=~36~K, with both temperatures indicated as vertical dotted lines; a linear trend is fitted to the temperature dependence between $T_N$ and $T^*$. (b) Integrated intensity of the $0.6 < |Q| < 0.9 \mathrm{\AA^{-1}}$ inelastic  scattering in Fig. 4, integrated in energy over $2 < E < 3.5$ meV, showing the formation of a gapped state as $T$ drops below $T_N$. }
 \end{figure}

Fig. 5(a) shows the magnetic order parameter measurement, taken from Gaussian fits to the [100] magnetic Bragg peak in Fig. 2(a).  For reference, both $T_N$~=~36~K and $T^*$ = 47~K are shown as vertical dashed lines.  The temperature dependence of the in-gap inelastic magnetic scattering can be further quantified by integrating the scattering in Fig. 4 in energy, over 2 $<$ $E$ $<$ 3.5~meV, as shown in Fig. 5(b).    We see a strong correlation between the temperature dependence of the order parameter and the in-gap magnetic scattering.  The order parameter in Fig. 5(a) shows two temperature regimes: a conventional downwards curvature regime below $T_N$, and a linear regime between $T_N$ and $T^*$.  Previous susceptibility and heat capacity measurements\cite{tomoko_3_2} have observed distinct signatures at both $T_N$ and $T^*$, suggesting that both temperature scales are relevant to this system.  The inelastic scattering within the gap shows an inflection point in its temperature dependence near $T_N$, but the gap begins to form at temperatures as high as $T^*$.  As elastic magnetic Bragg peaks are observed at all temperatures below $T^*$, we conclude that two distinct low-temperature regimes exist, with 3-D long-range order existing all the way up to $T^*$, but the fully-formed gap of $\Delta$ = 5~meV exists only below $T_N$. 

 \begin{figure}[h]
 \includegraphics[width=85mm]{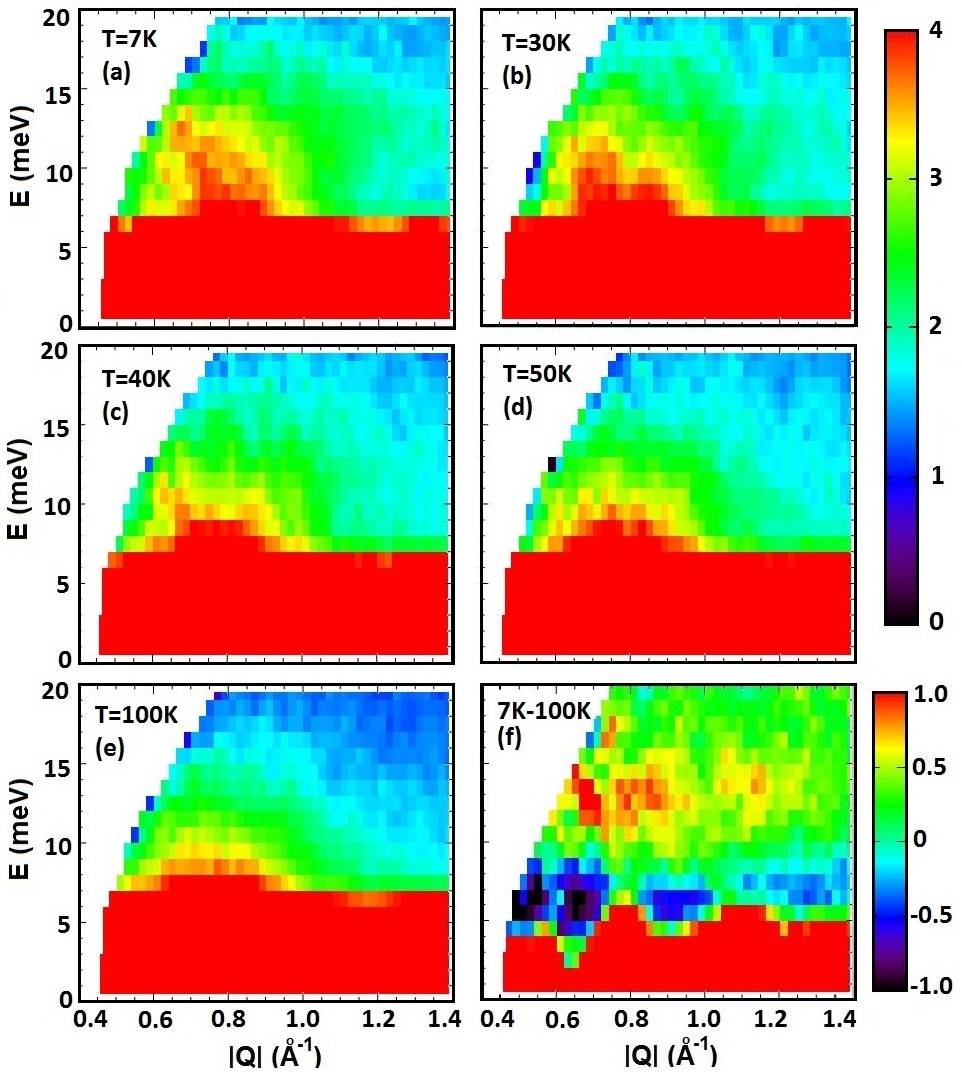}
 \caption{\label{}(a)-(e) Background-subtracted scattering is shown as a function of $|Q|$ and $E$ at five temperatures below and above $T_N$, using E$_i$=120 meV neutrons.   (f) Difference in scattered inelastic intensity between $T$~=~7~K and $T$~=~100~K; here the top of the magnetic scattering band can be seen at $E$~=~14~meV.  The upper intensity scale refers to panels (a)-(e), whereas the lower intensity scale refers to panel (f).}
 \end{figure}

 \begin{figure}[h]
 \includegraphics[width=85mm]{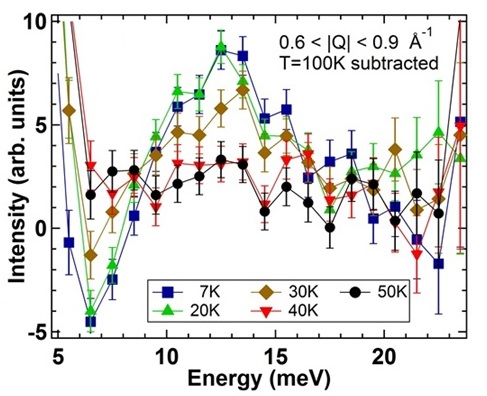}
 \caption{\label{}  Energy cuts of difference in scattered intensity ($T-$100~K) in the 0.6 $< |Q| < $ 0.9 $\mathrm{\AA}^{-1}$ wavevector band are shown.   Here a low-temperature excess of scattering is seen for energies near the top of the magnetic scattering band at $E$~=~14~meV.}
 \end{figure}

Measurements using $E_i$ = 120 meV allow a larger bandwidth to the spin excitations to be probed, and these are shown in Fig. 6.  In Figs. 6(a)$-$(e), the empty-can-subtracted scattering intensity is plotted at five temperatures between $T$~=~7~K and $T$~=~100~K.  In Fig. 6(f), the difference in scattering intensity between 7~K and 100~K is shown.   The magnetic Bragg peaks near the elastic channel, as well as the low-temperature deficiency of the scattering in the in-gap regime, can be seen.    The excess low-temperature spin wave scattering extends up to about 14~meV.   

In order to better quantify the energy dependence of the scattering intensities in Fig. 6, Fig. 7 shows cuts through different temperature data sets, integrated in $|Q|$ from 0.6 $\mathrm{\AA}^{-1}$ to 0.9~$\mathrm{\AA}^{-1}$, all subtracting $T$ = 100~K data sets in the same way as is shown for $T$ = 7~K in Fig. 6(c), for temperatures from 7~K to 50~K.  These data sets show clearly that the top of the spin excitation band is $\sim$14~meV, and the spin excitation spectrum has fully softened to its high temperature form by $\sim$40~K.

\section{DISCUSSION AND CONCLUSIONS}

We note that the spin gap that develops below $T_N$ is remarkably large, $\Delta$ $\sim$ 5~meV (comparable to the long-range ordering temperature $T^*$ = 47~K), and is almost half the spin-wave bandwidth [$\Delta$/(14 meV$-\Delta$)~$\sim$~1/2].  Our new results beg the question of the origin of the large spin gap.  As already mentioned, in the absence of SOC, little or no spin gap is expected as orbital angular momentum will be quenched.  However, when SOC is significant, a $d^3$ state corresponds not to a half-occupied and orbitally-quenched $t_{2g}$ triplet, but instead to a partially-filled $j_{eff} = 3/2$ quartet.  In systems with magnetic 3$d$ electrons, SOC is generally expected to be weak relative to typical orbital splitting energy scales, but in 5$d$ electronic systems such as iridates and osmates SOC splitting is often the dominant energy scale, as $Z^4$ is $\sim$50-150$\times$ larger in the latter.   4$d$ electronic systems such as Ba$_2$YRuO$_6$ may represent a crossover regime, in which SOC effects are significant, but not as dominant as in the analogous 5$d$ cases, leading to ground states characterized by physics relevant to both the non-SOC (orbitally quenched $t_{2g}^3$ triplet) and strong-SOC ($j_{eff}$~=~3/2) limits.  We consequently attribute the large $\Delta$~$\sim$~5~meV spin gap to moderately strong SOC appropriate to the 4$d$ magnetic electrons in Ru$^{5+}$.  

Comparison to the 4$d^1$ candidate spin liquid system Ba$_2$YMoO$_6$ is interesting as it also displays a very large spin gap, $\Delta$ $\sim$ 28 meV, but with a relatively narrow $\sim$4~meV bandwidth.  The top of the spin excitation spectrum in Ba$_2$YMoO$_6$ is a factor of 2 higher than that in Ba$_2$YRuO$_6$, which is consistent with a factor of 2 difference between the exchange constants $J$ estimated from high temperature susceptibility, with $\Theta_{CW}$=$Jn$s(s+1)/3, where s=$^1$/$_2$ for 4$d^1$ Mo$^{5+}$, s=$^3$/$_2$ for 4$d^3$ Ru$^{5+}$, and $n$ is the number of nearest magnetic neighbors.  This assumes an appropriate high-temperature susceptibility such that the SOC splitting is smaller than the high temperatures employed in the analysis of the susceptibility.  We see that the large gap in the spin excitation spectrum of Ba$_2$YRuO$_6$ is a defining physical characteristic of the low-temperature properties of this material, and is likely responsible for the two distinct ordered states it displays below $T^*$ and $T_N$.  As we infer that this gap arises from anisotropy expected in the limit of strong SOC, we conclude that the unusual and intriguing phase behavior of Ba$_2$YRuO$_6$ is a characteristic of its geometrically frustrated FCC ground state in the presence of SOC.

\begin{acknowledgments}
Work at McMaster University was supported by NSERC.  Work at Villanova University was sponsored by a Faculty Development Grant.  Research at Oak Ridge National Laboratory's Spallation Neutron Source was sponsored by the Scientific User Facilities Division, Office of Basic Energy Sciences, U. S. Department of Energy. 
\end{acknowledgments}

%\bibliography{BYRO}


\end{document}